# DETERMINING MINIMUM NUMBER OF REQUIRED ACCELEROMETER FOR OUTPUT-ONLY STRUCTURAL IDENTIFICATION OF FRAMES

*M.R. Davoodi[1], B. Navayi neya[2], S.A. Mostafavian[3], S.R. Nabavian[4], GH.R. Jahangiry[5]*

[1] Associate Professor of Civil Engineering, Babol University of Technology, Iran, davoodi@nit.ac.ir

[2] Associate Professor of Civil Engineering, Babol University of Technology, Iran, navayi@nit.ac.ir

[3] Assistant Professor of Civil Engineering, Payame Nor University, Sari branch, Iran, a.mostafavian@stu.nit.ac.ir

[4] Ph.D Student of Civil Engineering, Faculty Member, Tabari University of Babol, Iran, nabavian@stu.nit.ac.ir

[5] M.sc. of Structural Engineering, Tabari University of Babol, Iran, gr.jahangiry5011@gmail.com

## ABSTRACT

Operational modal analysis (OMA) aims at identifying the modal properties of a structure based on response data of the structure excited by ambient sources. Modal parameters of the ambient vibration structures consist of natural frequencies, mode shapes and modal damping ratios. In this paper, a typical frame with arbitrary loading has been modeled in finite element software, ANSYS, and the responses (accelerations of nodes) have been achieved. By using these data and the codes written in MATLAB, the modal parameters (natural frequencies, mode shapes) of the beams are identified with FDD (frequency Domain Decomposition) and PP (Peak Picking) methods and then justified with the results of input-output method which was determined by modal analysis. The results indicate that there is a good agreement between three methods for determining dynamic characteristics of frames. Then, minimum number of required accelerometer has been determined that recording the acceleration in that location lead to identify all the natural frequencies of frame.

## 1. INTRODUCTION

Identification of the global dynamic properties of civil engineering structures using vibration responses is a necessary step in several types of analysis including, for instance, model updating and structural health monitoring (SHM) [1]. The first step for SHM is system identification. During the last thirty years, civil engineers began to take advantage of a number of techniques developed in the system identification and experimental modal analysis field. Such techniques allowed the experimental identification of dynamic properties of structures by applying input-output modal identification procedures called Experimental Modal Analysis (EMA) [2]. Because of many limitation about using EMA for real structures [3], attention has been paid by the civil engineering community to Operational Modal Analysis (OMA) with applications on several structures. Operational modal analysis (OMA) has been used as a technology for estimating modal parameters of structures using output-only data, i.e., without the knowledge of input excitation [4, 5]. Modal parameters of the ambient vibration structures consist of natural frequencies, mode shapes and modal damping ratios. So far, a number of mathematical models on the output-only identification techniques have been developed and roughly classified by either parametric methods in the time domain or nonparametric ones in the frequency domain. Each identification method in either the time domain or the frequency one has its own advantage and limitation. If a model is fitted to data, the technique is parametric [6].

These techniques are more complex and computationally demanding, and they usually perform better than the faster and easier non-parametric techniques, which, however, are preferred for initial insight into the identification problem. Generally, parametric methods such as Ibrahim time domain (ITD), Eigensystem realization algorithm (ERA) or Random decrement technique (RDT) are preferable for estimating modal damping but difficulty in natural frequencies, mode shapes extraction, whereas nonparametric ones such as Peak-picking (PP), Frequency domain decomposition (FDD) or Enhanced frequency domain decomposition (EFDD) advantage on natural frequencies, mode shapes extraction, but uncertainty in damping estimation [7], [8]. Recently, new approach based on Wavelet transform (WT) and Hilbert-Huang transform (HHT) have been developed for output-only identification techniques in the time-frequency plane. The most undemanding method for output-only modal parameter identification is the basic frequency domain (BFD) technique [9], also called the Peak-Picking method, because the identification of eigenfrequencies is based on peak picking in the power spectrum plots. However, this method can lead to erroneous results if the basic assumptions of low damping and well-separated frequencies are not fulfilled. In fact, the method identifies the operational deflection shapes, which are the superpositions of multiple modes for closely spaced modes. The singular value decomposition (SVD) of the power spectral density (PSD) matrix overcomes these shortcomings and leads to the frequency domain decomposition (FDD) method [10], which is capable of detecting mode-multiplicity. However, both of these techniques are non-parametric methods because the modal parameters are obtained without fitting a mathematical model to the measured data.

In this article, the ambient vibration of a typical frame under ambient is studied and with accelerations data in all nodes of frame, the natural frequencies and the mode shapes are determined using PP and FDD methods. Then, the natural frequencies of frames are identified using only one acceleration data (recording the acceleration just in one location).

## 2. PICK PEAKING

The peak-picking method (PP) is the simpler and more practical method for modal identification. In spite of some drawbacks, the PP method gives very fast results and is useful as a pre-process tool when dynamic monitoring is performed. The PP method was systematized by Felber [11]. In this method the natural frequencies of the structures are determined as the peaks of the Average Normalized Power-Spectral Densities (ANPSDs) [12], the damping factors are determined using the Half Power Bandwidth Method [13], and the components of the mode shapes are determined by the values of the transfer functions at the natural frequencies [14]. The main limitations of the PP method, is that picking the peaks is often a subjective task, identifying close frequencies is difficult, spurious modes can be confused as real ones, operational deflection shapes are obtained instead of mode shapes and the damping estimates are unreliable [14].

### 2.1. BACKGROUND THEORY OF PP:

The displacements of the structure can be expressed as linear combination of the mode shapes:

$$\{x(t)\} = \{\emptyset_1\}y_1(t) + \{\emptyset_2\}y_2(t) + \cdots + \{\emptyset_n\}y_n(t)$$

Where $\emptyset_j$ is mode shape vector and $y_j$ is normalized coordinates. For convenient manipulation, you can transfer this equation into the frequency domain to yield:

$$\{X(\omega)\} = \{\emptyset_1\}Y_1(\omega) + \{\emptyset_2\}Y_2(\omega) + \cdots + \{\emptyset_n\}Y_n(\omega)$$

Where the normalized coordinates $Y_i(\omega)$ are defined as:

$$Y_j(\omega) = H_j(\omega)P_j(\omega)$$

Where $P_j(\omega)$ is the nodal excitation and the frequency response function of the jth mode $H_j(\omega)$ is given by:

$$H_j(\omega) = \frac{1}{K_j - \omega^2 M_j + i\omega C_j}$$

Using the fact $\ddot{X}(\omega) = \omega^2 X(\omega)$, you can express the accelerations of the structure in the frequency domain as:

$$\{\ddot{X}(\omega)\} = \omega^2[\{\emptyset_1\}H_1(\omega)P_1(\omega) + \{\emptyset_2\}H_2(\omega)P_2(\omega) + \cdots + \{\emptyset_n\}H_n(\omega)P_n(\omega)]$$

Or:

$$\ddot{X}_i(\omega) = \omega^2[A_{1i}H_1(\omega) + A_{2i}H_2(\omega) + \cdots + A_{ni}H_n(\omega)]$$

Where $A_{ji} = \emptyset_{ji}P_j(\omega)$

Setting $A_{ji} = 1$ for all js (for white noise), you can compute the magnitude of the individual acceleration response functions $\omega^2|H_j(\omega)|$ and their combination $|\ddot{X}_i(\omega)|$. By plotting the combined response $|\ddot{X}_i(\omega)|$, you can clearly see peaks corresponding to the damped frequencies of system. This indicates the natural frequencies of a structure can be estimated using Fourier transforms of ambient vibration acceleration records. If two acceleration records obtained simultaneously at location a and b are used, you can estimate the modal amplitude ratio of the jth mode for the two locations using:

$$\frac{\ddot{X}_a(\omega_j)}{\ddot{X}_b(\omega_j)} \cong \frac{\emptyset_{ja}\omega_j^2 H_j(\omega_j)P_j(\omega_j)}{\emptyset_{jb}\omega_j^2 H_j(\omega_j)P_j(\omega_j)} = \frac{\emptyset_{ja}}{\emptyset_{jb}}$$

You can then determine the mode shapes experimentally using the ratios of the Fourier amplitudes at natural frequencies.

## 2-2- ESTIMATING NATURAL FREQUENCIES:

Natural frequencies $\omega_j$ of structures are determined from the peaks of the auto spectra or power spectral density (PSD) of an ambient vibration acceleration record. This PSD is defined as

$$G_{ii}(\omega) = \ddot{X}_i(\omega)\ddot{X}_i^*(\omega)$$

$G_{ii}(\omega)$ corresponds to the square of the magnitude of the complex valued acceleration response and $\ddot{X}_i^*(\omega)$ is the complex conjugate of $\ddot{X}_i(\omega)$. PSD has local peaks where $\omega = \omega_d$ and, for small damping values, $\omega_d \cong \omega_n$.

Felber introduced a new function instead of PSD function in order to determine the frequencies. This function called the Averaged Normalized Power Spectral Density (ANPSD) is defined as the average of a group of $l$ normalized power spectral densities (NPSDs). The ANPSD functions are calculated using:

$$\text{ANPSD}(f_k) = \frac{1}{l}\sum_{i=1}^{i=l} NPSD_i(f_k)$$

Where $NPSD_i(f_k)$ is defined as

$$NPSD_i(f_k) = \frac{PSD_i(f_k)}{\sum_{k=0}^{k=n} PSD_i(f_k)}$$

And $f_k$ is the kth discrete frequency and n is the number of discrete frequencies.

## 2-3- ESTIMATING MODE SHAPES:

The mode shapes estimates $\emptyset_j$ associated with the natural frequency estimates $\omega_j$ are determined using a series of simultaneously recorded ambient vibration time histories $X_j(t)$. You can compute the transfer functions $T_{il}(\omega)$ between the Fourier transform of individual signals using:

$$T_{il}(\omega) = \frac{\ddot{X}_l(\omega)}{\ddot{X}_i(\omega)} = \frac{\ddot{X}_l(\omega)\ddot{X}_i^*(\omega)}{\ddot{X}_i(\omega)\ddot{X}_i^*(\omega)}$$

This complex valued transfer function can also be expressed in terms of its magnitude $|T_{il}(\omega)|$ and phase angle $\theta_{il}(\omega)$. You can estimate the absolute value of the ratio $|r_{j_{il}}| = \left|\frac{\phi_{jl}}{\phi_{ji}}\right|$ of the jth modal amplitudes between coordinates $i$ and $l$ using:

$$|r_{j_{il}}| \cong |T_{il}(\omega_j)|$$

The sign of the modal ratio, $r_{j_{il}}$, is determined from the phase angle $\theta_{il}(\omega_j)$. When the phase angle is near $0^o$, movements of the degrees of freedom $i$ and $l$ are in phase and the modal ratio is positive. On the other hand, when $\theta_{il}(\omega_j)$ is near $180^o$, the two motions of the coordinates are perfectly out of phase and the modal ratio is negative.

## 3. FREQUENCY DOMAIN DECOMPOSITION

Among the nonparametric methods in the frequency domain, FDD has been very widely used recently for output-only system identification through the ambient vibration measurements due to its reliability, straightforward and effectiveness [15], applied for wind-excited structures [8]. FDD is also powerful for closed natural frequencies extraction. However, FDD always requires the prior-selected natural frequencies as well as its strict hypotheses of uncorrelated white noise excitations and lightly structural damping. Under these strict hypotheses, the output PSD matrix can be expressed similarly as form of conventionally matrix decompositions, consequently, first-order linear approximation of the output PSD matrix is used for estimating mode shapes and damping. Output PSD matrix can be decomposed via fast-decayed decomposition methods such as QR decomposition and almost Singular value decomposition (SVD) [16]. By doing so, the spectral densities functions are decomposed in the contributions of different modes of a system that, at each frequency, contribute to its response. From the analysis of the singular values it is possible to identify the auto power spectral density functions corresponding to each mode of a system. In the FDD method, the mode shapes are estimated as the singular vectors at the peak of each auto power spectral density function corresponding to each mode. The FDD technique is an extension of the BFD method [17]. The theoretical basis can be summarized as follows. The relationship between the input x(t) and the output y(t) can be written in the following form [10]:

$$[G_{yy}(\omega)] = [H(\omega)]^*[G_{xx}(\omega)][H(\omega)]^T$$

Where $G_{xx}(\omega)$ is the r × r input PSD matrix; r is the number of inputs; $G_{yy}(\omega)$ is the m × m output PSD matrix; m is the number of outputs; $[H(\omega)]$ is the m × r FRF matrix; and the superscripts * and T denote complex conjugate and transpose, respectively. The FRF matrix can be expressed in a typical partial fraction form, which is used in classical modal analysis, in terms of poles, l, and residues, [R]:

$$[H(\omega)] = \frac{[Y(\omega)]}{[X(\omega)]} = \sum_{k=1}^{n}\left(\frac{[R_k]}{j\omega - \lambda_k} + \frac{[R_k]^*}{j\omega - \lambda_k^*}\right)$$

with

$$\lambda_k = -\sigma_k + j\omega_{dk}$$

Where n is the number of modes; $\lambda_k$ is the pole of the kth mode; $\sigma_k$ is the modal damping decay constant; and $\omega_{dk}$ is the damped natural frequency of the kth mode. [Rk] is the residue, and it is given by

$$[R_k] = \{\phi_k\}\{\gamma_k\}^T$$

Where $\{\phi_k\}$ is the mode shape vector, and $\{\gamma_k\}$ is the modal participation vector. Therefore, combining eq. (1) and (2) and assuming that the input is random in both time and space and has a zero mean white noise distribution (i.e., the PSD is constant: $[G_{xx}(\omega)] = [C]$), the output PSD matrix can be written as

$$[G_{yy}(\omega)] = \sum_{k=1}^{n} \sum_{s=1}^{n} \left[ \frac{[R_k]}{j\omega - \lambda_k} + \frac{[R_k]^*}{j\omega - \lambda_k^*} \right] [C] \left[ \frac{[R_k]}{j\omega - \lambda_k} + \frac{[R_k]^*}{j\omega - \lambda_k^*} \right]^H$$

Using the Heaviside partial fraction theorem for polynomial expansions, the following result can be obtained:

$$[G_{yy}(\omega)] = \sum_{k=1}^{n} \left( \frac{[A_k]}{j\omega - \lambda_k} + \frac{[A_k]^*}{j\omega - \lambda_k^*} + \frac{[B_k]}{-j\omega - \lambda_k} + \frac{[B_k]^*}{-j\omega - \lambda_k^*} \right)$$

This is the pole-residue form of the output PSD matrix. [Ak] is the kth residue matrix of the output PSD; it is an m × m hermitian matrix given by

$$[A_k] = [R_k][C] \sum_{s=1}^{n} \left( \frac{[R_s]^H}{-\lambda_k - \lambda_s^*} + \frac{[R_s]^T}{-\lambda_k - \lambda_s} \right)$$

If only the kth mode is considered, the following contribution is obtained:

$$[A_k] = \frac{[R_k][C][R_k]^H}{2\sigma_k}$$

This term can become dominant if the damping is low, and a residue proportional to the mode shape vector can be obtained as follows:

$$[A_k] \propto [R_k][C][R_k]^H = \{\phi_k\}\{\gamma_k\}^T [C] \{\gamma_k\}\{\phi_k\}^T = d_k \{\phi_k\}\{\phi_k\}^T$$

Where dk is a scaling factor for the kth mode. For a lightly damped system in which the contribution of the modes at a particular frequency is limited to a finite number (usually one or two), the response spectral density matrix can be written in the following final form:

$$[G_{yy}(\omega)] = \sum_{k \in Sub(\omega)} \left( \frac{d_k \{\phi_k\}\{\phi_k\}^T}{j\omega - \lambda_k} + \frac{d_k^* \{\phi_k\}^*\{\phi_k\}^{*T}}{j\omega - \lambda_k^*} \right)$$

Where $k \in Sub(\omega)$ is the set of contributing modes at the considered frequency. The SVD of the output PSD matrix known at discrete frequencies $\omega = \omega_i$ gives

$$[\hat{G}_{yy}(j\omega_i)] = [U]_i [S]_i [U]_i^H$$

Where the matrix [U]i is a unitary matrix holding the singular vector {uij}, and [S]i is a diagonal matrix holding the scalar singular values sij. Near a peak corresponding to the kth mode in the spectrum, this mode will be dominant. If only the kth mode is dominant, only one term in eq. (10) exists, and the PSD matrix approximates to a rank one matrix:

$$\hat{G}_{yy}(j\omega_i) = s_i \{u_{i1}\}\{u_{i1}\}^H \quad \omega_i \to \omega_k$$

In such a case, therefore, the first singular vector {ui1} represents an estimate of the mode shape:

$$\{\hat{\phi}\} = \{u_{i1}\}$$

And the corresponding singular value belongs to the auto power spectral density function of the SDOF system corresponding to the mode of interest. In the case of repeated modes, the PSD matrix rank is equal to the multiplicity number of the modes. The auto power spectral density function of the corresponding SDOF system is identified around the peak of the singular value plot by comparing the mode shape estimate $\{\hat{\phi}\}$ with the singular vectors associated with the frequency lines around the peak [18].

## 4. Numerical FE Model

First, a typical frame was modeled in finite element software, ANSYS, And the first five mode shapes has been achieved using modal analysis (Fig. 1).

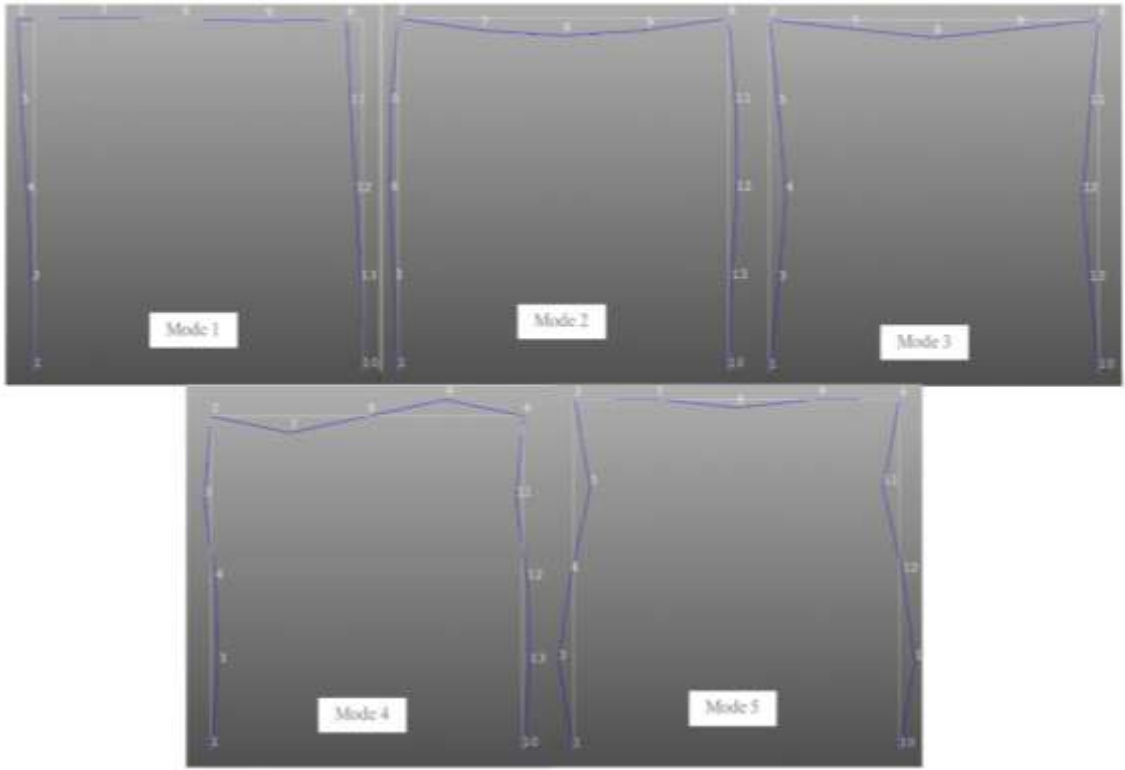

**Fig. 1- First five mode shapes of the frame**

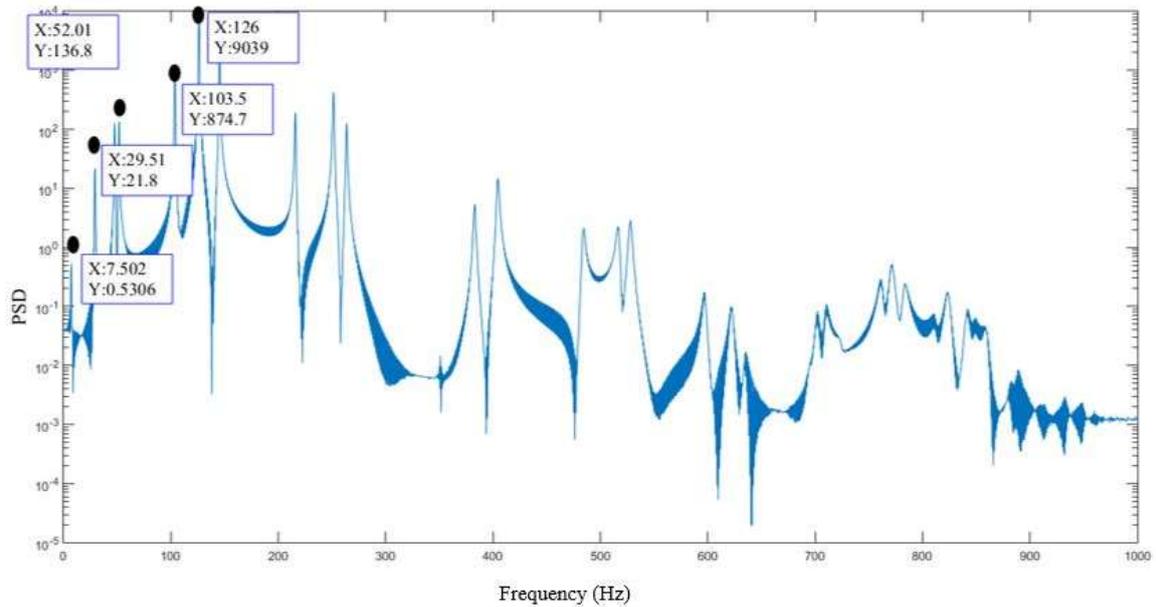

**Fig. 2- PSD plot of the Frame for x direction**

Then the frame was subjected to arbitrary ambient vibration and the acceleration response of frame has been obtained. The PSD diagram for the acceleration response of the frame is drawn with PP and FDD methods (Fig. 2-5).

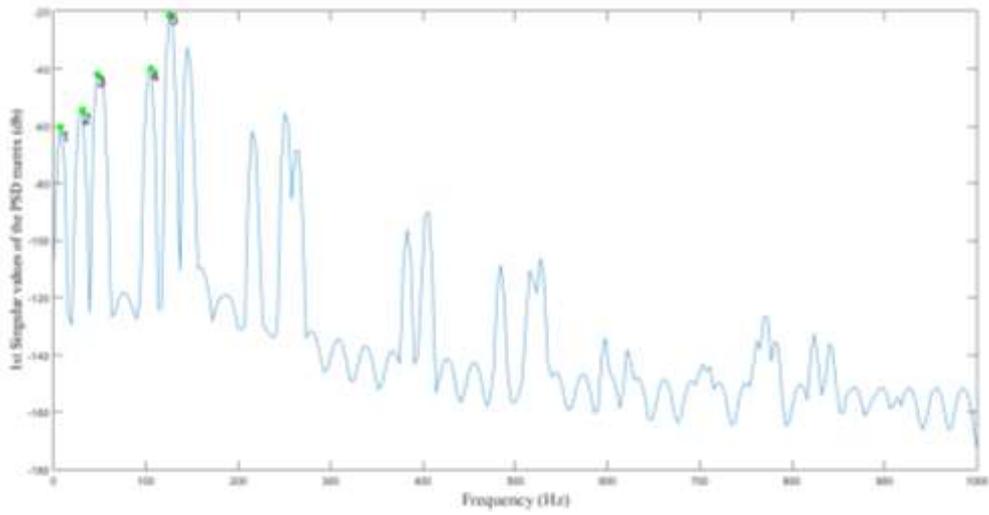

**Fig. 3- PSD plot of the Frame for x direction (FDD)**

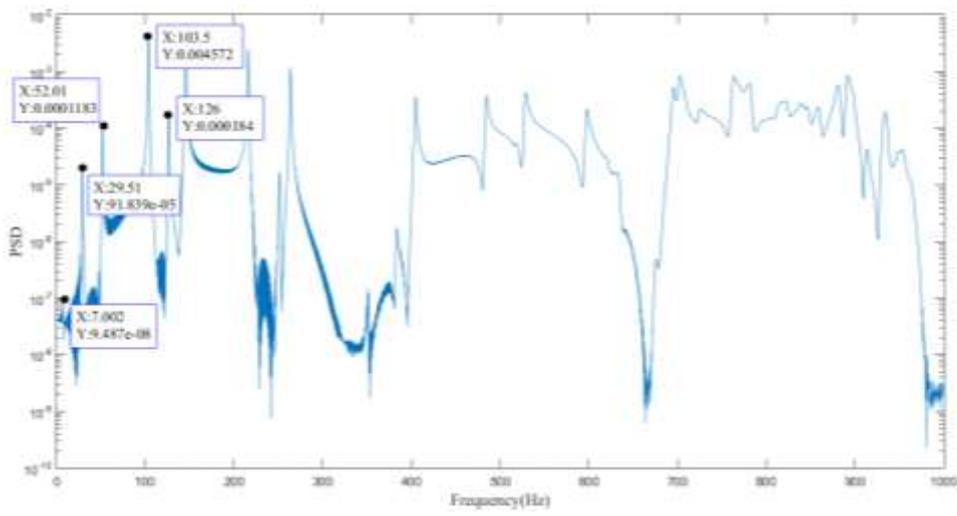

**Fig. 4- PSD plot of the Frame for y direction**

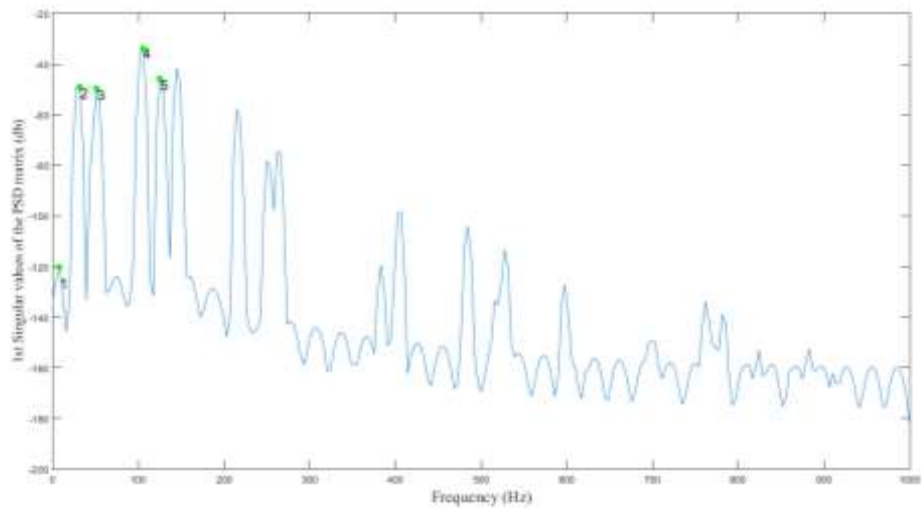

**Fig. 5- PSD plot of the Frame for y direction (FDD)**

In the tables 1-2, the mode shape values (vertical and horizontal deformation of the frame) for the frame with 12 elements has been shown with FEM, PP and FDD methods. According to the tables, there is a good agreement between the mode shapes in different methods.

**Table 1. Mode shape values for X direction**

| Node / Method | 1 | 3 | 4 | 5 | 2 | 7 | 8 | 9 | 6 | 11 | 12 | 13 | 10 |
|---|---|---|---|---|---|---|---|---|---|---|---|---|---|
| **Mode 1** | | | | | | | | | | | | | |
| FEM | 0 | 0.13622 | 0.44300 | 0.77379 | 1 | 1 | 1 | 1 | 1 | 0.77379 | 0.44300 | 0.13622 | 0 |
| PP | 0 | 0.13750 | 0.44638 | 0.77553 | 0.99996 | 1 | 0.99987 | 0.99994 | 0.99991 | 0.76868 | 0.45999 | 0.16417 | 0 |
| FDD | 0 | 0.14322 | 0.44718 | 0.77169 | 1 | 1 | 1 | 1 | 1 | 0.76963 | 0.43209 | 0.12995 | 0 |
| **Mode 2** | | | | | | | | | | | | | |
| FEM | 0 | 0.42472 | 1 | 0.93596 | 0.00011 | 0.00006 | 0 | -0.00006 | -0.00011 | -0.93596 | -1 | -0.42472 | 0 |
| PP | 0 | 0.42475 | 1 | 0.93663 | 0.02507 | 0.02504 | 0.02503 | -0.02502 | -0.02498 | -0.93212 | -0.97666 | -0.40036 | 0 |
| FDD | 0 | 0.42887 | 0.99463 | 0.92434 | 0.00606 | 0.00596 | 0.00596 | -0.00596 | -0.00596 | -0.92442 | -1 | -0.43186 | 0 |
| **Mode 3** | | | | | | | | | | | | | |
| FEM | 0 | 0.52849 | 1 | 0.52577 | -0.38787 | -0.38800 | -0.38805 | -0.38800 | -0.38787 | 0.52577 | 1 | 0.52849 | 0 |
| PP | 0 | 0.50953 | 0.96240 | 0.50426 | -0.37839 | -0.37850 | -0.37854 | -0.37850 | -0.37834 | 0.53545 | 1 | 0.52217 | 0 |
| FDD | 0 | 0.48433 | 0.91830 | 0.48172 | -0.37245 | -0.37245 | -0.37245 | -0.37245 | -0.37226 | 0.52603 | 1 | 0.52908 | 0 |
| **Mode 4** | | | | | | | | | | | | | |
| FEM | 0 | 1 | 0.63386 | -0.87739 | 0.14915 | 0.14939 | 0.14948 | 0.14939 | 0.14915 | -0.87739 | 0.63386 | 1 | 0 |
| PP | 0 | 1 | 0.63925 | -0.86554 | 0.15312 | 0.15337 | 0.15347 | 0.15340 | 0.15314 | -0.86924 | 0.61838 | 0.98886 | 0 |
| FDD | 0 | 0.99937 | 0.63194 | -0.87703 | 0.14913 | 0.14934 | 0.14955 | 0.14934 | 0.14913 | -0.87724 | 0.63299 | 1 | 0 |
| **Mode 5** | | | | | | | | | | | | | |
| FEM | 0 | -0.94000 | -0.21000 | 1 | 0.00217 | 0.00109 | 0 | -0.00108 | -0.00217 | -1 | 0.21000 | 0.94000 | 0 |
| PP | 0 | -0.93794 | -0.20900 | 0.99765 | 0.00458 | 0.00394 | 0.00351 | -0.00339 | -0.00359 | -1 | 0.20966 | 0.93988 | 0 |
| FDD | 0 | -0.93915 | -0.21001 | 0.99902 | 0.00239 | 0.00124 | 0.00044 | -0.00106 | -0.00200 | -1 | 0.20981 | 0.94014 | 0 |

**Table 2. Mode shape values for Y direction**

| Node / Method | 1 | 3 | 4 | 5 | 2 | 7 | 8 | 9 | 6 | 11 | 12 | 13 | 10 |
|---|---|---|---|---|---|---|---|---|---|---|---|---|---|
| **Mode 1** | | | | | | | | | | | | | |
| FEM | 0 | -0.00026 | -0.00052 | -0.00078 | -0.00104 | 1 | 0 | -1 | 0.00105 | 0.00078 | 0.00052 | 0.00026 | 0 |
| PP | 0 | -0.00015 | -0.00036 | -0.00049 | -0.00062 | 1 | 0.33901 | -0.69188 | 0.00097 | 0.00078 | 0.00050 | 0.00027 | 0 |
| FDD | 0 | -0.00013 | -0.00013 | -0.00026 | -0.00026 | 1 | 0.15467 | -0.84936 | 0.00066 | 0.00053 | 0.00040 | 0.00013 | 0 |
| **Mode 2** | | | | | | | | | | | | | |
| FEM | 0 | 0.00010 | 0.00021 | 0.00031 | 0.00041 | 0.68045 | 1 | 0.68045 | 0.00041 | 0.00031 | 0.00021 | 0.00010 | 0 |
| PP | 0 | 0.00010 | 0.00020 | 0.00030 | 0.00040 | 0.67994 | 1 | 0.68383 | 0.00041 | 0.00031 | 0.00021 | 0.00010 | 0 |
| FDD | 0 | 0.00014 | 0.00028 | 0.00028 | 0.00042 | 0.67960 | 1 | 0.67794 | 0.00042 | 0.00028 | 0.00014 | 0.00014 | 0 |
| **Mode 3** | | | | | | | | | | | | | |
| FEM | 0 | 0.00027 | 0.00055 | 0.00082 | 0.00110 | 0.54398 | 1 | 0.54398 | 0.00110 | 0.00082 | 0.00055 | 0.00027 | 0 |
| PP | 0 | 0.00028 | 0.00056 | 0.00084 | 0.00111 | 0.53337 | 1 | 0.55762 | 0.00110 | 0.00082 | 0.00055 | 0.00027 | 0 |
| FDD | 0 | 0.00025 | 0.00051 | 0.00076 | 0.00114 | 0.57886 | 1 | 0.509401 | 0.00114 | 0.00088 | 0.00051 | 0.00020 | 0 |
| **Mode 4** | | | | | | | | | | | | | |
| FEM | 0 | -0.00075 | -0.00150 | -0.00226 | -0.00300 | -1 | 0 | 1 | 0.00300 | 0.00226 | 0.00151 | 0.00075 | 0 |
| PP | 0 | -0.00077 | -0.00153 | -0.00230 | -0.00304 | -0.99622 | 0.01022 | 1 | 0.00299 | 0.00224 | 0.00150 | 0.00075 | 0 |
| FDD | 0 | -0.00071 | -0.00156 | -0.00226 | -0.00297 | -1 | 0.00040 | 1 | 0.00297 | 0.00226 | 0.00156 | 0.00071 | 0 |
| **Mode 5** | | | | | | | | | | | | | |
| FEM | 0 | 0.00025 | 0.00050 | 0.00075 | 0.00100 | -0.14171 | 1 | -0.14171 | 0.00100 | 0.00075 | 0.00050 | 0.00025 | 0 |
| PP | 0 | 0.00026 | 0.00053 | 0.00079 | 0.00104 | -0.15018 | 1 | -0.15004 | 0.00102 | 0.00076 | 0.00051 | 0.00025 | 0 |
| FDD | 0 | 0.00020 | 0.00051 | 0.00071 | 0.00102 | -0.14281 | 1 | -0.14142 | 0.00102 | 0.00071 | 0.00051 | 0.00020 | 0 |

The following charts are related to MAC diagram for FEM and PP mode shapes and FEM and FDD mode shapes (Fig. 6-7).

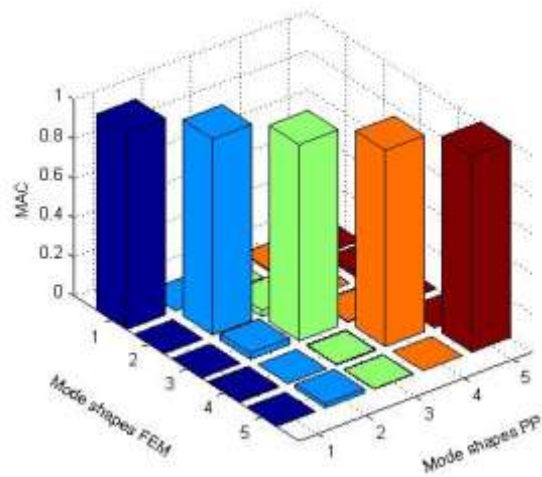

**Fig. 6- MAC diagram for mode shapes (FEM and PP)**

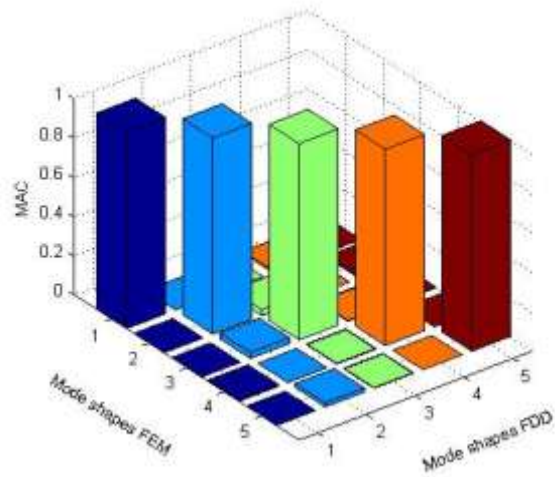

**Fig. 7- MAC diagram for mode shapes (FEM and FDD)**

In the following table (tables 3-4) the values of the natural frequencies of FEM, PP and FDD are compared with together and the corresponding graph is drawn (Fig. 8-9).

**Table 3. Natural Frequencies values for frame x direction**

| Method | Frequency 1 | Frequency 2 | Frequency 3 | Frequency 4 | Frequency 5 |
|---|---|---|---|---|---|
| FEM | 7.61 | 30.04 | 53.11 | 107.29 | 131.03 |
| PP | 7.502 | 29.01 | 52.01 | 104 | 126 |
| FDD | 7.81 | 31.25 | 50.78 | 105.5 | 125 |

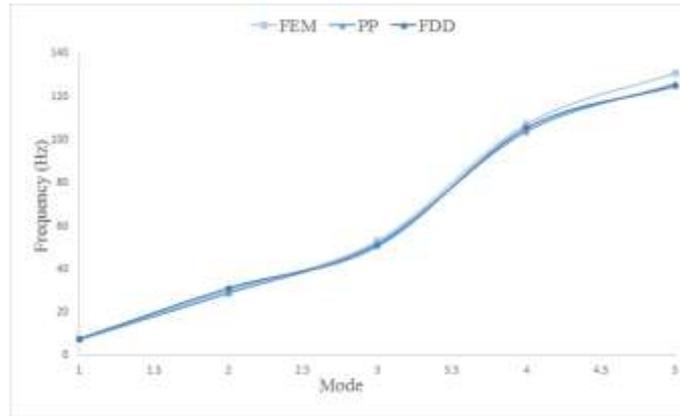

**Fig. 8- Natural Frequencies for X direction**

**Table 4. Natural Frequencies values for Y direction**

| Method | Frequency 1 | Frequency 2 | Frequency 3 | Frequency 4 | Frequency 5 |
|---|---|---|---|---|---|
| FEM | 7.61 | 30.04 | 53.11 | 107.29 | 131.03 |
| PP | 7.002 | 29.51 | 51.51 | 103.5 | 126 |
| FDD | 7.81 | 31.25 | 50.78 | 105.5 | 125 |

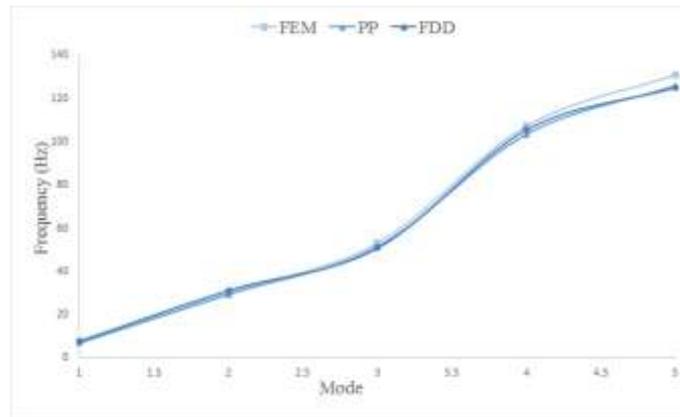

**Fig. 9- Natural Frequencies for Y direction**

The following tables (tables 5-6) refer to the identified natural frequencies of the frame by measuring the acceleration in just one place.

**Table 5. Frequency values for X direction by 1 Accelerometer**

| Node / Mode | 3 | 4 | 5 | 2 | 7 | 8 | 9 | 6 | 11 | 12 | 13 |
|---|---|---|---|---|---|---|---|---|---|---|---|
| 1st Mode | 7.502 | 7.502 | 7.502 | 7.502 | 7.502 | 7.502 | 7.502 | 7.502 | 7.502 | 7.502 | 7.502 |
| 2nd Mode | 29.51 | 29.51 | 29.01 | - | - | - | - | - | - | 29.51 | 29.51 |
| 3rd Mode | 47.51 | 47.51 | 48.01 | 47.51 | 47.51 | 47.51 | 47.51 | 47.51 | 47.51 | 48.01 | 48.01 |
| 4th Mode | 103.5 | 103.5 | 104 | 103.5 | 103.5 | 103.5 | 103.5 | 103.5 | 104 | 104 | 104 |
| 5th Mode | 126 | 126 | 126 | 126.5 | 126.5 | 126 | 126 | 126 | 126 | 126 | 126 |

**Table 6. Frequency values for Y direction by 1 Accelerometer**

| Node / Mode | 3 | 4 | 5 | 2 | 7 | 8 | 9 | 6 | 11 | 12 | 13 |
|---|---|---|---|---|---|---|---|---|---|---|---|
| 1st Mode | 7.002 | 7.002 | 7.002 | 7.002 | 7.502 | - | 8.002 | 7.502 | 7.502 | 7.502 | 7.002 |
| 2nd Mode | 29.01 | 29.51 | 29.51 | 29.51 | 29.01 | 29.51 | 29.01 | 29.51 | 29.01 | 29.51 | 29.51 |
| 3rd Mode | - | - | - | - | 48.01 | - | 47.51 | - | - | - | - |
| 4th Mode | 103.5 | 103.5 | 104 | 103.5 | 103.5 | - | 103.5 | 104 | 104 | 103.5 | 104 |
| 5th Mode | 126 | 126 | 126 | 126 | 126.5 | 126 | 126 | 126.5 | 126.5 | 126.5 | 126.5 |

As can be seen, if the acceleration data in x direction is used, by placing an accelerometer on each of the Nodes 3, 4, 5, 12 and 13, all natural frequencies of the frame is detected. And if the acceleration data in y direction is used, by placing an accelerometer on each of the Nodes 7 and 9, all natural frequencies of the frame is detected.

## 5. Results:

In this paper, output-only identification methods are used to estimate the modal parameters of a two dimensional frame from a record of output measurements. The performance of two different system identification methods have been compared and discussed. The two methods are the frequency domain based peak-picking methods and FDD (Frequency Domain Decomposition). The comparison reveals that the two methods give reasonable estimates of the natural frequencies and mode shapes that are in consistent with those obtained from input-output modal analysis of the frame. And also, it has been shown that there are some locations in the frame where by recording acceleration data on that location, it possible to identify natural frequencies of the frame